\documentstyle[12pt]{article}
\begin{document}
\thispagestyle{empty}
\newcommand{\be}{\begin{equation}}
\newcommand{\ee}{\end{equation}}
\newcommand{\sect}[1]{\setcounter{equation}{0}\section{#1}}
\newcommand{\vs}[1]{\rule[- #1 mm]{0mm}{#1 mm}}
\newcommand{\hs}[1]{\hspace{#1mm}}
\newcommand{\mb}[1]{\hs{5}\mbox{#1}\hs{5}}
\newcommand{\bea}{\begin{eqnarray}}
\newcommand{\eea}{\end{eqnarray}}
\newcommand{\wt}[1]{\widetilde{#1}}
\newcommand{\ux}[1]{\underline{#1}}
\newcommand{\ov}[1]{\overline{#1}}
\newcommand{\sm}[2]{\frac{\mbox{\footnotesize #1}\vs{-2}}
           {\vs{-2}\mbox{\footnotesize #2}}}
\newcommand{\prt}{\partial}
\newcommand{\eps}{\epsilon}\newcommand{\p}[1]{(\ref{#1})}
\newcommand{\R}{\mbox{\rule{0.2mm}{2.8mm}\hspace{-1.5mm} R}}
\newcommand{\Z}{Z\hspace{-2mm}Z}
\newcommand{\cd}{{\cal D}}
\newcommand{\cg}{{\cal G}}
\newcommand{\ck}{{\cal K}}
\newcommand{\cw}{{\cal W}}
\newcommand{\vj}{\vec{J}}
\newcommand{\vl}{\vec{\lambda}}
\newcommand{\vz}{\vec{\sigma}}
\newcommand{\vt}{\vec{\tau}}
\newcommand{\poiss}{\stackrel{\otimes}{,}}
\newcommand{\tx}{\theta_{12}}
\newcommand{\tb}{\overline{\theta}_{12}}
\newcommand{\zw}{{1\over z_{12}}}
\newcommand{\sqp}{{(1 + i\sqrt{3})\over 2}}
\newcommand{\sqm}{{(1 - i\sqrt{3})\over 2}}
% REVUES POUR BIBLIO
\newcommand{\NP}[1]{Nucl.\ Phys.\ {\bf #1}}
\newcommand{\PLB}[1]{Phys.\ Lett.\ {B \bf #1}}
\newcommand{\PLA}[1]{Phys.\ Lett.\ {A \bf #1}}
\newcommand{\NC}[1]{Nuovo Cimento {\bf #1}}
\newcommand{\CMP}[1]{Commun.\ Math.\ Phys.\ {\bf #1}}
\newcommand{\PR}[1]{Phys.\ Rev.\ {\bf #1}}
\newcommand{\PRL}[1]{Phys.\ Rev.\ Lett.\ {\bf #1}}
\newcommand{\MPL}[1]{Mod.\ Phys.\ Lett.\ {\bf #1}}
\newcommand{\BLMS}[1]{Bull.\ London Math.\ Soc.\ {\bf #1}}
\newcommand{\IJMP}[1]{Int.\ J.\ Mod.\ Phys.\ {\bf #1}}
\newcommand{\JMP}[1]{Jour.\ Math.\ Phys.\ {\bf #1}}
\newcommand{\LMP}[1]{Lett.\ Math.\ Phys.\ {\bf #1}}
\renewcommand{\thefootnote}{\fnsymbol{footnote}}
\newpage
\setcounter{page}{0}
\pagestyle{empty}
\vs{12}
\begin{center}
{\LARGE {\bf Triality of Majorana-Weyl Spacetimes}}\\
{\LARGE {\bf with Different Signatures}}\\
[0.8cm]

\vs{10} {\large M.A. De Andrade$^{(a,b)}$, M. Rojas$^{(a)}$ and F.
Toppan$^{(a,c)}$} ~\\ \quad \\
 {\em {~$^{(a)}$} CBPF, DCP, Rua Dr. Xavier Sigaud
150, cep 22290-180 Rio de Janeiro (RJ), Brazil}\\
~\quad \\
 {\em {~$~^{(b)}$} UCP, FT, Rua
Bar{\~{a}}o do Amazonas, 124, cep 25685-070, Petr\'{o}polis (RJ),
Brazil}\\ ~\quad\\ {\em ~$~^{(c)}$ UFES, CCE Depto de
F{\'{\i}}sica, Goiabeiras cep 29060-900, Vit\'oria (ES), Brazil}

\end{center}
\vs{6}

\centerline{ {\bf Abstract}}

\vs{6}

Majorana-Weyl spacetimes offer a rich algebraic setup and new types
of space-time dualities besides those discussed by Hull. The triality
automorphisms
of $Spin(8)$ act non-trivially on Majorana-Weyl representations and
Majorana-Weyl spacetimes with different signatures.
In particular relations exist among the
$(1+9)\leftrightarrow (5+5)\leftrightarrow (9+1)$ spacetimes,
as well as their transverse coordinates
spacetimes $(0+8)\leftrightarrow (4+4)\leftrightarrow (8+0)$.
Larger dimensional spacetimes such as
$(2+10)\leftrightarrow (6+6)\leftrightarrow (10+2)$
also show dualities induced by
triality. A precise three-languages dictionary is here given.
It furnishes the exact translations among, e.g., the three
different versions (one in each signature) of the ten-dimensional $N=1$
superstring and superYang-Mills theories. Their dualities close
the six-element permutation group $S_3$. Bilinear and trilinear
invariants allowing to formulate theories with a manifest
space-time symmetry are constructed.

\vs{6} \vfill \rightline{July 1999} \rightline{CBPF-NF-039/99}
\rightline{hep-th/xxx} {\em E-Mails:\\ 1) marco@cbpf.br\\ 2)
mrojas@cbpf.br \\
3) toppan@cbpf.br}
\newpage
\pagestyle{plain}
\renewcommand{\thefootnote}{\arabic{footnote}}
\setcounter{footnote}{0}
\vs{8}

\section{Introduction.}

Physical theories formulated in different-than-usual spacetimes
signatures have recently found increased attention. One of the
reasons can be traced to the conjectured $F$-theory \cite{Vafa}
which supposedly lives in $(2+10)$ dimensions \cite{Nish}.
The current interest in AdS theories motivated by the AdS/CFT
correspondence furnishes another motivation. Two-time physics
e.g. has started been explored by Bars and collaborators in a
series of papers \cite{Bars}. For another motivation we can also
recall that a fundamental theory is
expected to explain not only the spacetime dimensionality, but
even its signature (see \cite{Duff}). Quite recently Hull and
Hull-Khuri \cite{Hull} pointed out the existence of dualities
relating different compactifications of theories formulated in
different signatures. Such a result provides new insights to the
whole question of spacetime signatures.
Other papers (the most recent is \cite{Cori}) have remarked the
existence of space-time dualities.\par
Majorana-Weyl spacetimes (i.e. those supporting Majorana-Weyl
spinors) are at the very core of the present knowledge of
the unification via supersymmetry, being at the basis of
ten-dimensional superstrings, superYang-Mills and
supergravity theories (and perhaps the already mentioned
$F$-theory). A well-established feature of Majorana-Weyl
spacetimes is that they are endorsed of a rich structure.
A legitimate question one can ask oneself is whether they are
affected, and how, by space-time dualities. The answer is quite
surprising (in fact it should not be so, with afterthought),
the structure of dualities is much richer than expected and
potentially able to shed a complete new light on the subject.
Indeed all different Majorana-Weyl spacetimes which are possibly
present in any given dimension are each-other related by duality
transformations which are induced by the $Spin(8)$ triality
automorphisms.
The action of the triality automorphisms is quite non-trivial and
has far richer consequences than the ${\bf Z}_2$-duality
(its most trivial representative) associated to the space-time
$(s,t)\leftrightarrow (t,s)$ exchange discussed in \cite{Duff}.
It corresponds to $S_3$, the six-element group of permutations
of three letters, identified with the group of congruences of the
triangle and generated by two reflections.
The lowest-dimension in which the triality action is non-trivial
is $8$ (not quite a coincidence), where the spacetimes
$(8+0)-(4+4)-(0+8)$ are all interrelated. They correspond to the
transverse coordinates of the $(9+1)-(5+5)-(1+9)$ spacetimes
respectively, where the triality action can also be lifted. Triality
relates as well the $12$-dimensional Majorana-Weyl spacetimes
$(10+2)-(6+6)-(2+10)$, i.e. the potentially interesting cases for
the $F$-theory, and so on.\\
As a consequence of triality, supersymmetric theories formulated
with Majorana-Weyl spinors in a given dimension but with different
signatures, are all dually mapped one into another.
A three-language dictionary is here furnished with the exact
translations among, e.g., the different versions of the $10$-dimensional
MW supersymmetric theories, formulated in the Majorana-Weyl
representation.\par
The strategy here followed is based in three steps. At first it
is shown that Majorana-Weyl spacetimes in dimensions $d> 8$
can be recovered from the properties of the $8$-dimensional
Majorana-Weyl
spacetimes and $\Gamma$-matrices representations. Next, working in
$d=8$, we construct, for each one of the three Majorana-Weyl
spacetimes $(8+0)$, $(4+4)$, $(0+8)$, the ``bridge transformations"
relating the corresponding Majorana-Weyl representations to the
representations (called ``VCA" in the text) which exhibit manifest
triality among vectors, chiral and antichiral spinors. As a final
step new ``bridge transformations" of spacetime kind, relating
among them the VCA representations constructed in each one
of the Majorana-Weyl spacetimes above, are given.\par
We emphasize that, contrary to Hull \cite{Hull}, the dualities
here discussed are already present for the {\it uncompactified}
theories and in this respect look more fundamental.\par
Moreover, bilinear and trilinear invariants under the $S_3$
permutation group of the three Majorana-Weyl spacetimes are
constructed. They can be possibly used to formulate supersymmetric
Majorana-Weyl theories in a manifestly triality-invariant form
which presents an explicit symmetry under exchange of space and
time coordinates.\par
The present paper is intended to be an abridged version, suitable
for a letter-size, of a forthcoming extended version which
presents in full detail the construction and where extra results
which are outside the scope of this letter are also furnished.\par
The scheme of this work is as follows. In the next section we
recall, following \cite{Kugo} and \cite{DeAn}, the basic properties
of
$\Gamma$-matrices and Majorana conditions needed for our
construction. Majorana-type representations are analyzed in
section $3$. We show how to relate the Majorana-Weyl representations
in $d> 8$ to the $8$-dimensional Majorana-Weyl representations.
In section $4$ we introduce, for $d=8$, the set of data necessary
to
define a supersymmetric Majorana-Weyl theory, i.e.
the set of ``words" of our three-languages dictionary. The Cartan's
\cite{Cart}
triality among vectors, chiral and antichiral spinors is presented
in section $5$. The main result is furnished in section $6$, where
spacetime triality is discussed. In the Conclusions we furnish some
comments and point out some perspectives.

\vspace{0.2cm}
\noindent{\section{Preliminary results.}}

Here we limit ourselves to introduce the basic ingredients needed
for our constructions. Further information is found in \cite{Kugo}
and \cite{DeAn}.\par
We denote as $g_{mn}$ the flat (pseudo-)euclidean metric of a
$(t+s)$-spacetime.
Time
(space) directions in our conventions are associated to the $+$
(respectively $-$) sign. \par
The $\Gamma$'s matrices are assumed to be unitary (the chosen normalization
is for the square of time-like 
$\Gamma$-matrices being $+1$). The three matrices
${\cal A}$, ${\cal B}$, ${\cal C}$ are the generators of the three
conjugation operations (hermitian, complex conjugation and transposition
respectively) on the $\Gamma$'s.
In particular
\begin{eqnarray}
{\cal C} \Gamma^m {\cal C}^\dagger &=& \eta(-1)^{t+1}{\Gamma^m}^T
\label{0}
\end{eqnarray}
where $\eta=\pm 1$ in even-dimensional spacetimes label
inequivalent choices of the charge conjugation matrix ${\cal
C}$.\par
${\cal A}, {\cal B}, {\cal C}$ are related by the formula
\begin{eqnarray}
{\cal C}&=& {\cal B}^T {\cal A} \label{1}
\end{eqnarray}
Up to an inessential phase, ${\cal A}$ is specified by the product of all the
time-like $\Gamma$ matrices.
An unitary transformations $U$ applied on spinors act on $\Gamma^{m}$,
${\cal A}, {\cal B}, {\cal C}$ according to \cite{DeAn}
\begin{eqnarray}
\Gamma^m &\mapsto& U\Gamma^m U^\dag\nonumber\\ {\cal A}&\mapsto&
U{\cal A}U^{\dag}\nonumber\\{\cal B}&\mapsto& U^\ast {\cal B}
U^{\dag}\nonumber\\
{\cal C}&\mapsto & U^\ast{\cal C} U^\dag \label{2}
\end{eqnarray}
A Majorana representation for the $\Gamma$'s can be defined as the one
in which ${\cal B}$ is set equal to the identity. Spinors can be
assumed real in this case.\par
In even dimensions we can also introduce the notion of Weyl
representation, i.e. when the ``generalized $\Gamma^5$ matrix" is
symmetric and block diagonal and with no loss of generality
can be assumed to be the direct sum of the two equal-size blocks
${\bf 1}\oplus (-{\bf 1})$.
The compatibility of both Majorana and Weyl conditions
constraints the spacetime $(t+s)$ to satisfy
\begin{eqnarray}
s-t&=& 0 \quad {\it mod}\quad 8, \quad {\it for} \quad{\it both}\quad
{\it values}
\quad \eta=\pm 1
\label{3}
\end{eqnarray}
In even dimensions
Majorana representations, but not of Weyl type, are also found for
\begin{eqnarray}
s-t &=& 2 \quad  mod \quad 8 \quad  for \quad \eta=-1;\nonumber \\
s-t &=& 6 \quad mod \quad 8 \quad for \quad \eta= +1.
\label{4}
\end{eqnarray}
For $d<8$ the only spacetimes supporting Majorana-Weyl spinors
have signatures $(n+n)$. At $d=8$ a new feature arises,
Majorana-Weyl spinors can be found for different signatures.

 \vspace{0.2cm}
\noindent{\section{Majorana-type representations.}}

It is convenient to introduce the notion of Majorana-type
representation (or shortly MTR) for the $\Gamma$ matrices as one in
which all the $\Gamma$'s have a definite symmetry. For $d= p+q$
a MTR with $p$ symmetric and $q$ antisymmetric $\Gamma$'s will
be denoted as
$(p_S, q_A)$ in the following. \par
For such representations the ${\cal C}$ matrix introduced in the
previous section is given by either the product of all the symmetric
$\Gamma$ matrices, denoted as ${\cal C}_S$, or all the antisymmetric
ones (${\cal C}_A$). In even dimensions ${\cal C}_S$,
${\cal C}_A$ correspond to opposite values of $\eta$ in (\ref{0}).\par
A Majorana representation in a given signature spacetime is a MTR.
Conversely, given a MTR, we can find a spacetime signature for
which the representation is Majorana. The admissible couples of
$(p_S, q_A)$ values for a MTR are immediately read from
the Majorana tables given above ({\ref{3}) and (\ref{4}). The
construction is such that ${\cal C}$ must correspond to the
correct value of $\eta$ in the tables.\par
The list of all possible MTR's in any given dimension is easily
computed. In order just to give an example one can check that in $d=6$
there exists a MTR (not of Weyl kind) with $6$ anticommuting $\Gamma$
matrices plus an anticommuting $\Gamma^7$ $(0_S, 6_A, {\Gamma^7}_A)$.
It gives the Majorana basis in the Euclidean $6$-dimensional space.
\par
In $d=8$ the MTR's of Weyl type are $(8_S,0_A)$,
$(4_S,4_A)$, $(0_S,8_A)$, associated to the corresponding Majorana-Weyl
spacetimes.\par
Different MTR's belong to different classes under similarity
transformations of the
$\Gamma$'s representations. Indeed
in, let's say, an euclidean (all $+$ signs) space, the
index
\begin{eqnarray}
&&I = tr (\Gamma^m \cdot {\Gamma_m}^T) = (p_S-q_A)\cdot tr{\bf 1}
\label{5}
\end{eqnarray}
takes different values for different MTR's.\par
We computed explicitly all MTR's up to
$d=12$ and Majorana-Weyl representations
up to $d=14$ (the results will be furnished elsewhere)
by using a recursive
algorithm presented in \cite{Cola}. It allows producing Weyl
representations in $d$ dimensions from any given couple of
representations in $r$ and $s$ dimensions, for even-dimensional $d,r,s$
satisfying $d=r+s+2$. The only MTR up to $d=12$ which does not directly
fit
into this scheme, the above-mentioned $6$-dimensional $(0_S,6_A)$,
is however constructed from the $(3_S,3_A)$ representation
(this one directly produced from the $2$-dimensional Pauli matrices for
$r=s=2$) after computing the value of the symmetric
matrix ${\cal
B}$ in the euclidean $6$-dimensional space, and later finding the
transformation (\ref{2}) which maps it into the unity.
\par
The algorithm is given by the formula
\begin{eqnarray}
{\Gamma_d}^{i=1,..., s+1} &=& \sigma_x\otimes {\bf 1}_L\otimes
{\Gamma_s}^{i=1,...,s+1}\nonumber\\
{\Gamma_d}^{s+1+j = s+2, ..., d} &=&
\sigma_y\otimes{\Gamma_r}^{j=1,...,r+1}\otimes {\bf 1}_R
\label{algo}
\end{eqnarray}
where ${\bf 1}_{L,R}$ are the unit-matrices in the respective
spaces, while $\sigma_x = e_{12}+e_{21}$ and $\sigma_y = -i e_{12}+i
e_{21}$ are the $2$-dimensional Pauli matrices.
${\Gamma_r}^{r+1}$ corresponds to the ``generalized $\Gamma^5$-matrix"
in $r+1$ dimensions. In the above
formula the values $r,s=0$ are allowed. The corresponding
${\Gamma_0}^1$ is just $1$.\par
With the help of the above formula we have a very efficient tool
to reduce the
analysis of Majorana-Weyl representations for $d\geq 10$ to the
$8$-dimensional case, since we can always set either $r$ or $s$
equal to $8$ (or possibly both, which corresponds to the $d=18$
case). Up to $d=14$ Majorana-Weyl spacetimes exist for three
different signatures and the same situation of $d=8$ is repeated.
A new
feature arises for $d\geq 16$. A careful analysis of the consistency
conditions is needed in this case, since the triality
transformations that we later discuss no longer preserve the similarity
classes of MTRs; stated otherwise, different representatives of MTR's
in the same similarity class and for the same couple of values
$(p_S,q_A)$ are mapped under a given triality transformation into
representatives of MTR belonging to different
similarity classes. This feature is likely to be related with the
problems encountered
in defining supersymmetric theories in dimensions greater
than $14$, which have been e.g. pointed out in \cite{Sezg}. \par
In any case the construction here discussed is suited to analyze and
works perfectly well for the range $d=8,...,14$, i.e. the cases which are
of interest
for, among the others, the superstrings and the F-theory (notice that
the above scheme can find applications to dualities also for odd-dimensional
Majorana spacetimes as the $11$-dimensional ones supporting the $M$-theory,
we will comment more on that in the conclusions).
We postpone to the extended version of this paper the presentation
of the full
set of reconstruction formulas which make explicit the
construction of MW-spacetimes
in dimensions $d>8$ in terms of the $8$-dimensional ones. For the
purpose of this paper is sufficient to recall that such
reconstruction formulas can be given. In particular the higher
dimensional ${\cal C}$ charge conjugation matrices for $d>8$ can
be expressed in terms of the eight-dimensional matrices ${\cal
C}_8$.
\vspace{0.2cm}
\noindent{\section{The set of data for $d=8$.}}

In this section we present the set of data needed to specify a
Majorana-Weyl supersymmetric theory formulated in $8$-dimensions.
As we have stated in the previous section, this set of data
can be ``lifted" to define Majorana-Weyl supersymmetric theories
formulated in higher-dimensions. The results here furnished
therefore have a more general validity.\par
At first we recall that we have three different Majorana-Weyl
spacetimes $(8+0)-(4+4)-(0+8)$ and two choices for $\eta=\pm 1$,
so in total $3\times 2=6$ inequivalent theories (i.e. inequivalent
versions of some given supersymmetric theory) that can be
formulated in $d=8$.\par
Each one of these versions is characterized
by the following set of data {\em all expressed in the
corresponding Majorana-Weyl representation}, being this one the most
suitable for analyzing supersymmetry.
Such data will play the roles of the
``words" in the three-languages dictionary that will be later
furnished:
\par
{\em i)} the bosonic (and/or vector-fields) coordinates $x_m$, with
vector index $m=1,...,8$;\par
{\em ii)} the fermionic coordinates (and/or spinorial fields)
$\psi_a$, $\chi_{\dot{a}}$, with chiral and antichiral
indices $a=1,...,8$, $\dot{a}=1,...,8$ respectively;\par
{\em iii)} the diagonal (pseudo-)orthogonal spacetime metric
$(g^{-1})^{mn}$, $g_{mn}$;\par
{\em iv)} the ${\cal A}$ matrix of section $2$, used to introduce barred
spinors, which is now decomposed in an equal-size block diagonal form
such as ${\cal A} = A\oplus {\tilde A}$, with structure of indices
${(A)_a}^b$ and ${(\tilde{A})_{\dot{a}}}^{\dot{b}}$
respectively;\par
{\em v)} the charge-conjugation matrix ${\cal C}$, always symmetric,
also put in
equal-size block diagonal form ${\cal C} = {C^{-1}}\oplus {\tilde
C}^{-1}$. Since ${\cal C}$ is invariant under bispinorial
transformations it can be promoted to be a metric for the space
of chiral (and respectively antichiral) spinors, used to raise and
lower spinorial indices. Indeed we can set
$(C^{-1})^{ab}$, $(C)_{ab}$, and $({\tilde C}^{-1})^{\dot{a}\dot{b}}$,
$({\tilde C})_{\dot{a}\dot{b}}$;\par
{\em vi) } finally we have the upper-left $\sigma$ and the lower-right
${\tilde \sigma}$ blocks in the $\Gamma$'s matrices with structure of indices
${(\sigma_m)_a}^{\dot{b}}$ and
${(\tilde{\sigma}_m)_{\dot{a}}}^b$ respectively.\par
The ${\cal B}$ matrix is automatically set to be the 
identity (${\cal B} ={\bf 1}$)
due to our choice of working in the Majorana-Weyl
representation.\par
In order to work in the Majorana-Weyl basis for each
one of the six different versions of the theory, the correct
Majorana-type representation must be picked up. In $(4+4)$ the
$(4_S,4_A)$ representation must be chosen for both values of
$\eta$, while in $(8+0)$ the $(8_S,0_A)$ works for $\eta=+1$ and
the $(0_S,8_A)$ works for $\eta=-1$ (and conversely in the $(0+8)$
case).\par
For later purpose it is convenient to present the matrix ${\cal
C}$ for each one of the six versions.
We have
\begin{eqnarray}
{\cal C} &=& \Gamma^9 = {\bf 1}_8\oplus (-{\bf 1}_8) ~\quad\quad\quad
\quad\quad in \quad
(8+0)\nonumber\\
{\cal C} &=& {\bf 1}_{16}\quad\quad\quad\quad\quad\quad\quad
\quad\quad\quad\quad
 in \quad
(0+8)\nonumber\\
{\cal C} &=& (C^{-1})\oplus (\eta C^{-1})\quad\quad\quad\quad
\quad ~~ in \quad (4+4)
\label{ccc}
\end{eqnarray}
where in the last case $C^{-1}$ can be chosen to be the $8\times
8$ matrix with $(+--++--+)$ entries in the antidiagonal and
$0$ entries in any other position.

\vspace{0.2cm}
\noindent{\section{The V-C-A triality.}}

The outer automorphisms of the $D_4$ Lie algebra is responsible
for the triality property among the $8$-dimensional vector, chiral
and antichiral spinor representations of $SO(8)$ which has been
first discussed by Cartan \cite{Cart}. \par
In our language the triality property can be restated as follows.
The $C^{-1}$, ${\tilde C}^{-1}$ matrices introduced in the
previous section, for each one of the six different cases we
discussed, are symmetric and with the same set (up to
an overall sign) of eigenvalues
($\pm 1$) as the corresponding $g^{-1}$ spacetime metric.\par
In each one of the above cases one can simultaneously map
both $C^{-1}$, ${\tilde
C}^{-1}$ (with different similarity transformations ${\cal G}$,
${\tilde {\cal G}}$) into the corresponding spacetime $g^{-1}$
metric:
\begin{eqnarray}
{\cal G}: \quad C^{-1} &\mapsto & g^{-1} = G\cdot C^{-1}\cdot
G^T\nonumber\\
{\tilde{{\cal G}}} : \quad {\tilde C}^{-1} &\mapsto & g^{-1} =
{\tilde G}\cdot {\tilde C}^{-1}\cdot {\tilde G}^T
\end{eqnarray}
The structure of indices for $G$ and its inverse $G^{-1}$ is
${(G)^m}_a$ and ${(G^{-1})^a}_m $ (an analogous structure holds for
${\tilde G}$, and ${\tilde G}^{-1}$). Therefore $G$, ${\tilde G}$
can be used to transform chiral (antichiral) indices in vector
indices allowing to work with, let's say, vector indices
alone.\par
The concrete $8\times 8$-dimensional matrices $G$, ${\tilde G}$ are
of course not uniquely defined since any other matrix of the kind
$L_C\cdot G\cdot L_g$, with $L_C$, $L_g$ preserving by
similarity the $C^{-1}$ and the $g^{-1}$ metrics respectively, are
equally well suited. For our purposes however it is sufficient to
furnish a concrete representative for the $G$, ${\tilde G}$
matrices.\par
In all the above cases we can choose the
concrete matrices $G$, ${\tilde G}$ to be
square root of unity:
\begin{eqnarray}
G^2 = {\tilde G}^2& =&{\bf 1}_8
\end{eqnarray}
Indeed the only case in the previous section discussion
where $C^{-1}$ (${\tilde
C}^{-1}$) is not diagonal is the $(4+4)$ case, with
$C^{-1}$ given in (\ref{ccc}) and ${\tilde C}^{-1} =\pm C^{-1}$.
A matrix $G$ which sets
$C^{-1}$ to the diagonal form $(++++----)$ is
\begin{equation}
G = \frac{1}{\sqrt{2}}\cdot
\left(
\begin{array}{cccccccc}
  1 & 0 & 0 & 0 & 0 & 0 & 0 & 1 \\
  0 & 1 & 0 & 0 & 0 & 0 &-1 & 0 \\
  0 & 0 & 1 & 0 & 0 &-1 & 0 & 0 \\
  0 & 0 & 0 & 1 & 1 & 0 & 0 & 0 \\
  0 & 0 & 0 & 1 &-1 & 0 & 0 & 0 \\
  0 & 0 &-1 & 0 & 0 &-1 & 0 & 0 \\
  0 &-1 & 0 & 0 & 0 & 0 &-1 & 0 \\
  1 & 0 & 0 & 0 & 0 & 0 & 0 &-1
\end{array}
\right)
\label{matrix}
\end{equation}
In all the other cases we have to flip a number
$n$ of signs in the diagonal, with $n=0$ $mod$ $4$.
Instead of working with the standard Wick rotation
prescription, which is the only one applicable
for {\em odd} numbers of signs
to be flipped, a smarter choice is allowed for {\em even} numbers
of flipping: the passage e.g. from $(++)\mapsto (--)$ can be
produced via similarity with the help of the $\sigma_y$ Pauli
matrix $\sigma_y = -i e_{12} +i e_{21}$, through
\begin{eqnarray}
\sigma_y \cdot{\bf 1}_2 \cdot {\sigma_y}^T &=& - {\bf 1}_2
\end{eqnarray}
Of course $\sigma_y$ satisfies ${\sigma_y}^2= {\bf 1}_2$
and is antisymmetric. The bridge matrices $G$ which flip the
euclidean $(++++++++)$ metric into the $(++++----)$ and
the $(--------)$
metrics are therefore respectively given by
\begin{eqnarray}
G_1 &=& {\bf 1}_4\oplus\sigma_y\oplus \sigma_y\nonumber\\
G_2 &=& \sigma_y\oplus\sigma_y\oplus\sigma_y\oplus\sigma_y
\label{bridges}
\end{eqnarray}
which are both square root of unity.\par
The above given bridge operators $G$, ${\tilde G}$ in a given
Majorana-Weyl spacetime allows to pass from the Majorana-Weyl
representation to another representation, that we can call VCA,
where triality is manifest and only vector-like indices are
present. Please notice that, as far as transformation properties
alone are concerned, the commuting or anticommuting nature of
spinors is not taken into account. For commuting spinors a more
radical property holds. Bilinear and trilinear invariants under
the $S_3$ permutation group of vectors, chiral and antichiral
spinors, can be constructed.
The procedure is as follows. At first the three bilinear scalars
\begin{eqnarray}
&&B_V = V^T \eta^{-1} V, \quad\quad
B_C = \psi^T C^{-1}\psi, \quad\quad
B_A = \chi^T {\tilde C}^{-1} \chi
\label{bilinear}
\end{eqnarray}
and the trilinear one
\begin{eqnarray}
T &=& \psi^T C^{-1}\sigma \eta^{-1}V\chi
\label{trilinear}
\end{eqnarray}
are constructed. Applying the above bridge transformations in a passive
way we can set
\begin{eqnarray}
{\hat \psi} = {(G^T)}^{-1} \psi,\quad &\quad&\quad
 {\hat \chi} = {({\tilde G}^T)}^{-1}\chi
\end{eqnarray}
Therefore
\begin{eqnarray}
&& B_V = V^T\eta^{-1}V, \quad\quad B_C = {\hat \psi}^T \eta^{-1}{\hat
\psi},\quad \quad B_A= {\hat \chi}^T\eta^{-1}{\hat\chi}
\end{eqnarray}
and their sum
\begin{eqnarray}
B&=& B_V+B_C+B_A
\end{eqnarray}
is by construction invariant under the $S_3$ exchange of $V,{\hat\psi}^T,
{\hat\chi}$.\par
As for the trilinear term $T$, it reads as follows
\begin{eqnarray}
T&=& {\hat\psi}^T M V {\hat \chi}
\end{eqnarray}
where the trivector $M$ is given by
\begin{eqnarray}
M^{mnp} &=& (\eta^{-1})^{mr} {({G^T}^{-1})_r}^a
{(\sigma_q)_a}^{\dot b} (\eta^{-1})^{qn}{{({\tilde G}^T)}_{\dot
b}}^p
\label{trivector}
\end{eqnarray}
A trilinear scalar, invariant under $S_3$, is  constructed through
\begin{eqnarray}
&& \sum_{perm} M^{mnp}V_m {{\hat \psi}^T}_n{\hat\chi}_p
\end{eqnarray}
where the sum is extended over all the permutations of $V,
{\hat\psi}^T, {\hat\chi}$.\par
The action of the $S_3$ permutation group on the original vectors,
chiral
and antichiral spinors is given by the pull-back of the bridge
transformations. It is just sufficient to write it down for two
of the generators, called $P$, $R$, of $S_3$, where
\begin{eqnarray}
P^2=R^2={\bf 1} \quad&\quad& \quad (PR)^3 ={\bf 1}
\end{eqnarray}
We have
\begin{eqnarray}
P: &&V\mapsto V, \quad\quad \quad {\psi^T} \leftrightarrow
{\tilde G}^T \chi G\nonumber\\
R: && \psi^T\mapsto\psi^T, \quad\quad V\leftrightarrow
{\tilde G}^T \chi
\end{eqnarray}

\vspace{0.2cm}
\noindent{\section{The spacetime triality.}}

The triality discussed in the previous section is the
Cartan's $V-C-A$ triality, which connects vectors, chiral and
antichiral spinors of the {\em same} spacetime. However
the procedure which has been used so far can be repeated to
connect vectors belonging to spacetime metrics with
different signatures.
In the case of our interest three spacetimes, denoted
as $X,Y,Z$, possess vector-indices
$(m, {\tilde m}, {\overline{m}}$)
which are referred to the metrics\\
$(++++++++)$, $(++++----)$,
$(--------)$
respectively.\par
The passage from one of the above
metrics to another one can be done by employing the same bridge
matrices introduced in (\ref{bridges}). The construction
straightforwardly repeat the one already encountered.
It should be clear that an enormous technical advantage is offered
by performing the connection between two different Majorana-Weyl
spacetimes working in both cases with the respective VCA
representations.
The ``spacetime bridge matrices" in this case only see
vector indices.
The connection between Majorana-Weyl representations
is then reconstructed in terms of the bridge matrices
(introduced in the previous section)
linking, in each one of the two spacetimes, the Majorana-Weyl
with the VCA representation. \par
There is no need to repeat here the formulas presented in section
5. Each one finds its ``mirror spacetime" equivalent. They have
just to be reinterpreted in the light of the spacetime
triality.\par
We just point out that under spacetime triality
$Y_{\tilde m}$, $Z_{\overline m}$ are mapped into
\begin{eqnarray}
Y_{\tilde m} \mapsto {\hat Y}_m,\quad&\quad&\quad
Z_{\overline m} \mapsto {\hat Z}_m
\end{eqnarray}
carrying a $(++++++++)$ vector index-structure.\par
The bilinear invariant under the $S_3$ group
of permutations is
\begin{eqnarray}
B&=& X^T{\eta_X}^{-1}X +{\hat Y}^T {\eta_X}^{-1}{\hat Y}
+{\hat Z}^T {\eta_X}^{-1}{\hat Z}
\end{eqnarray}
while the trilinear one is
\begin{eqnarray}
T&=& \sum_{perm} M^{mnp} X_m {{\hat Y}_n}^T {\hat Z}_p
\end{eqnarray}
where the trivector $M^{mnp}$ is constructed in full
analogy with (\ref{trivector}).\par
We remark that for what concerns spacetime triality invariances
it is likely that we do not
have to bother about the anticommuting character of spinors
as it is the case for invariances under $V-C-A$ triality.
For instance, in the simplest example, the bosonic
supersymmetric composite vector $X^m = x^m - i{\overline \theta}
\Gamma^m \theta $ is the building block to introduce the superparticle
and we do not need to worry about Grassmann
variables.
\par
Let us finally comment that the same bridge transformations
already
encountered can be used to produce dualities relating the opposite
values of $\eta$. Indeed just an overall $-$ sign for the
${\tilde C}^{-1}$ matrix distinguishes the two cases.

\vspace{0.2cm}
\noindent{\section{Conclusions.}}

In this paper we have shown that the triality automorphisms of
$Spin(8)$ not only
induce dualities among different-signatures Majorana-Weyl
spacetimes, but also furnish bilinear and trilinear invariants
which can be possibly used to formulate supersymmetric theories
possessing a space-time triality invariance. \par
Triality relations connect the various Majorana-Weyl spacetimes in
a given dimension.\par
The basic strategy of our construction consists in the fact that
even for dimensions $d>8$ the different signatures of Majorana-Weyl
spacetimes can be encoded in the $8$-dimensional $\Gamma$
matrices, used as building blocks in the construction of
the higher-dimensional $\Gamma$'s.\par
The bridge operators connecting in each given Majorana-Weyl
spacetime the Majorana-Weyl representation to a Cartan-type
representation in which the $8$-dimensional $V-C-A$ triality
is manifest are helpful in linking together different-signatures
Majorana-Weyl spacetimes. Indeed in a Cartan-like basis the
problem of relating different space-time signatures is
considerably simplified since we have to worry about just
how to connect vector-like indices and spacetime metrics.
\par
A complete and extensive list of the results here outlined will be
presented in a forthcoming paper.\par
The range of possible applications for the methods and the ideas here
discussed is vast. We limit ourselves to mention that we are currently
investigating the web of dualities connecting the $12$-dimensional
Majorana-Weyl spacetimes which should support the $F$-theory
($3\times 2$, taking into account of $\eta$), with the
$6$ versions of the $11$-dimensional Majorana spacetimes (for the
$M$-theory) in $(10+1)$, $(9+2)$, $(6+5)$, $(5+6)$, $(2+9)$,
$(1+10)$ signatures and with the different (again $3\times 2$)
versions of the $10$-dimensional Majorana-Weyl spacetimes.

\vskip1cm \noindent{\Large{\bf Acknowledgments}} \\ {\quad}\\ We
are pleased to acknowledge J. A. Helay\"{e}l-Neto and L.P. Colatto
for both encouragement and helpful discussions. We are grateful to
DCP-CBPF for the kind hospitality.

\end{document}